\pdfoutput=1
\documentclass[review]{elsarticle}

\usepackage{amssymb}
\usepackage{amsthm}
\usepackage{amsmath}
\usepackage{textcomp}
\usepackage[figuresright]{rotating}
\usepackage{lineno,hyperref}
\usepackage{subfig}

\usepackage{txfonts}

\usepackage[utf8]{inputenc}

\usepackage{lineno}

\usepackage{lipsum}

\usepackage{color}

\newcommand{\ee}{e$^+$e$^-$}

\begin{document}

\begin{frontmatter}



\title{The Heavy Photon Search beamline and its performance}


\newcommand{\red[1]}{{\color{red}{\bf #1}}}
\newcommand{\JLAB}{Thomas Jefferson National Accelerator Facility, Newport News, Virginia 23606}
\newcommand{\ODU}{Old Dominion University, Norfolk, Virginia 23529}
\newcommand{\GLASGOW}{University of Glasgow, Glasgow G12 8QQ, United Kingdom}
\newcommand{\UNH}{University of New Hampshire, Department of Physics, Durham, NH 03824}
\newcommand{\SLAC}{SLAC National Accelerator Laboratory, Menlo Park, CA 94025}

\author[JLAB]{ N. Baltzell}
\author[JLAB]{ H. Egiyan}
\author[ODU]{M. Ehrhart}
\author[SLAC]{C. Field}
\author[JLAB]{ A. Freyberger} 
\author[JLAB]{ F.-X. Girod} 
\author[UNH]{ M. Holtrop}
\author[SLAC]{ J. Jaros}
\author[ODU]{G. Kalicy}
\author[SLAC]{T. Maruyama\corref{corrauthor}}
\ead{tvm@slac.stanford.edu} 
\author[GLASGOW]{B. McKinnon}
\author[SLAC]{ K. Moffeit}
\author[SLAC]{T. Nelson} 
\author[SLAC]{ A. Odian} 
\author[SLAC]{ M. Oriunno} 
\author[UNH]{ R. Paremuzyan}
\author[JLAB]{ S. Stepanyan}
\author[JLAB]{M. Tiefenback}
\author[SLAC]{S. Uemura}
\author[JLAB]{ M. Ungaro} 
\author[ODU]{H. Vance}

\address[JLAB]{\JLAB}
\address[ODU]{\ODU}
\address[SLAC]{\SLAC}     
\address[UNH]{\UNH}
\address[GLASGOW]{\GLASGOW}

\cortext[corrauthor]{Corresponding author.}



\begin{abstract}
The Heavy Photon Search (HPS) is an experiment to search for a hidden sector photon, aka a heavy photon or 
dark photon, in fixed target electroproduction at
the Thomas Jefferson National Accelerator Facility (JLab). 
The HPS experiment searches for the e$^+$e$^-$ decay of the heavy photon with bump hunt and detached vertex strategies using a compact, 
large acceptance forward spectrometer, consisting of a silicon microstrip detector for tracking and vertexing, 
and a PbWO$_4$ electromagnetic calorimeter for energy measurement and fast triggering.
To achieve large acceptance and good vertexing resolution, the first layer of silicon detectors is placed  just $10$~cm 
downstream of the target with the sensor edges only $500~\muup$m above and below the beam. Placing the SVT in such close proximity to the beam puts stringent requirements on the beam profile and beam position stability. As part of an approved engineering run, HPS took data in 2015 and 2016 at $1.05$~GeV and $2.3$~GeV beam energies, respectively. 
This paper describes the beam line and its performance during that data taking.
\end{abstract}

\begin{keyword}
electron beam \sep collimator \sep heavy photon \sep silicon microstrips \sep electromagnetic calorimeter

\end{keyword}

\end{frontmatter}

\clearpage



\section{Introduction}
\label{introduction}

The Heavy Photon Search (HPS) experiment \cite{HPS_prop} at the Thomas Jefferson National Accelerator Facility is a search for a new $20-500$~MeV/c$^2$ vector gauge boson A$^\prime$ (``heavy photon'', aka ``dark photon" or ``hidden sector photon") in fixed target electro-production. Such a particle would couple weakly to electric charge by virtue of ``kinetic mixing"~\cite{kinetic_mixing}. Consequently A$^\prime$s could be produced by electron bremsstrahlung and decay to electron/positron pairs or pairs of other charged particles. Since the expected coupling is $\epsilon e$, with $\epsilon \le 10^{-3}$, A$^\prime$ production is small compared to standard QED production of \ee~pairs. To identify A$^\prime$s above this copious trident background, HPS looks for a sharp bump in \ee~ invariant mass and, for very small couplings, separated decay vertices. It does so with a compact 6-layer silicon vertex tracker (SVT) situated in a dipole magnetic field. A highly segmented PbWO$_4$ electromagnetic calorimeter (ECal) provides the trigger. Observing a statistically significant signal in the presence of a large background requires sufficient luminosity such that the number of signal events is large compared to the square root of the background (for the bump hunt), or a means of reducing the background so that it is negligible compared to the signal (for the vertex search).  This is accomplished with the continuous wave (CW) CEBAF beam, utilizing very fast electronics, a high rate trigger, and high data rate capability. To minimize the multiple scattering of the incident beam into the detector, HPS employs very thin target foils and relatively high ($\sim100$s~nA) average beam currents. HPS avoids most of these scattered beam electrons as well as those that have radiated and been bent in the horizontal plane by the dipole magnet, by splitting the SVT and ECal vertically into top and bottom sections, situated just above and below the beam.  The beam is passed through the entire apparatus in vacuum to minimize beam gas backgrounds.

The kinematics of the reaction are such that A$^\prime$s are produced at very forward angles with energy approximately that of the incident beam ($1-6$~GeV for HPS). For A$^\prime$ masses of interest, the A$^\prime$ decay products are very forward peaked, so the detectors must be placed as close to the beam as possible. Similarly, good vertex resolution requires the silicon tracker to be as close as possible to the target. The HPS detector accepts vertical scattering angles greater than $15$~mrad. The first silicon layer is positioned  $10$~cm downstream of the target, so the physical edge of the silicon sensor is placed just 500~$\muup$m above and below the beam (the sensor has a 1~mm wide guard ring). Proximity to the beam imposes stringent requirements on acceptable beam size, stability, and halo; necessitates protection collimation; and demands real time monitoring and circuitry to protect against errant beam motion.
The innermost silicon detectors see high but tolerable radiation levels and roughly 1~\% strip occupancies close to the beam. The innermost ECal crystals see roughly 1 MHz rates.

This paper will discuss HPS's beam requirements, the design of the HPS beamline, and its performance. It will review the beamline instrumentation used to measure and monitor performance and to protect against errant beam motion. The excellent quality and stability of the CEBAF beams coupled with HPS protection systems lets HPS take data safely with its silicon detectors just 500~$\muup$m from the  electron beam.

\section{HPS Beamline}
\label{beamlinedesign}

\begin{figure*}[t]
\begin{center}
\includegraphics[width=1.0\textwidth]{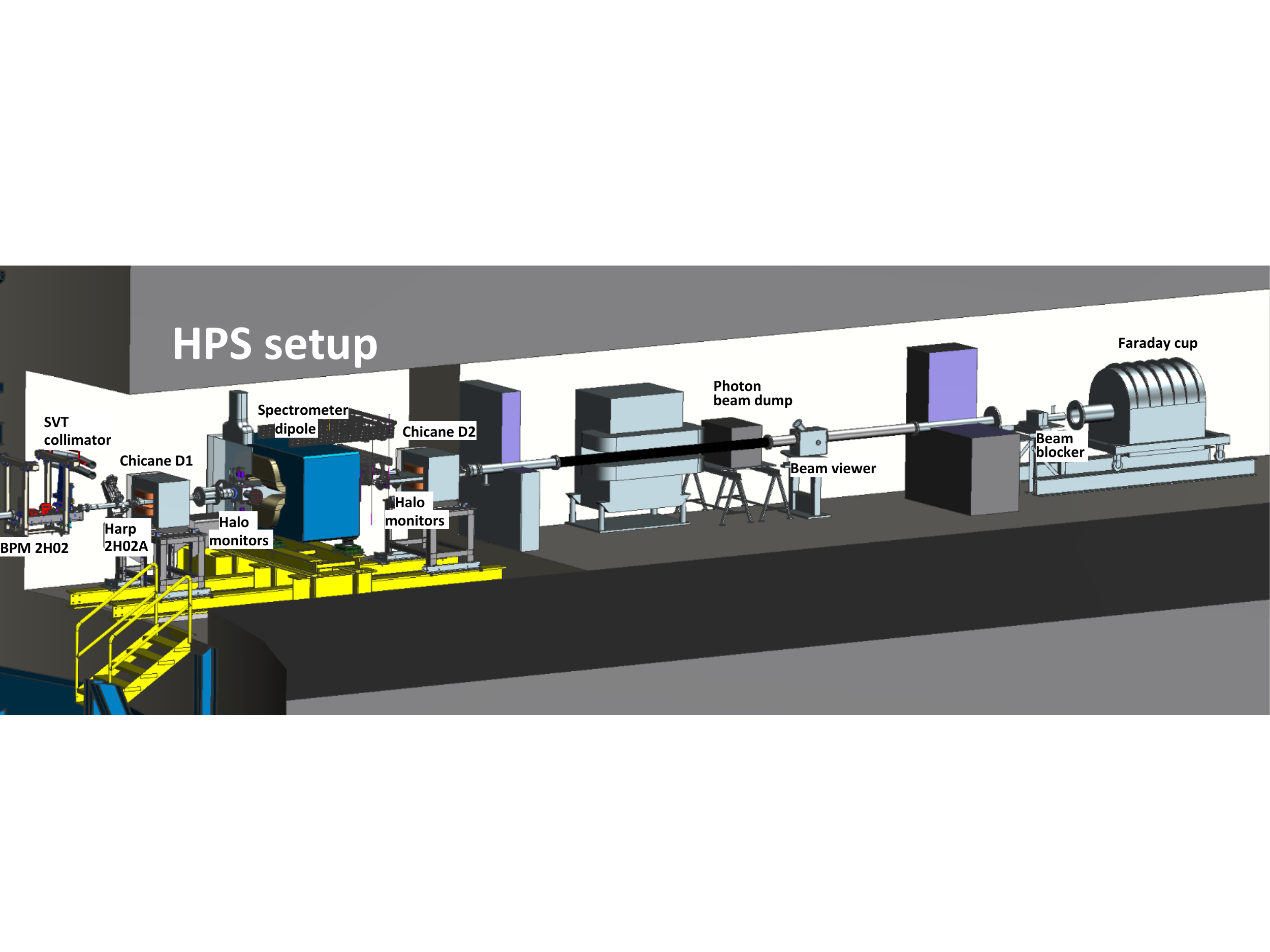}
\caption{The downstream end of the 2H beam line with the HPS setup.}
\label{fig:2h_line}
\end{center}
\end{figure*}

\begin{figure*}[t]
\begin{center}
\includegraphics[width=0.8\textwidth]{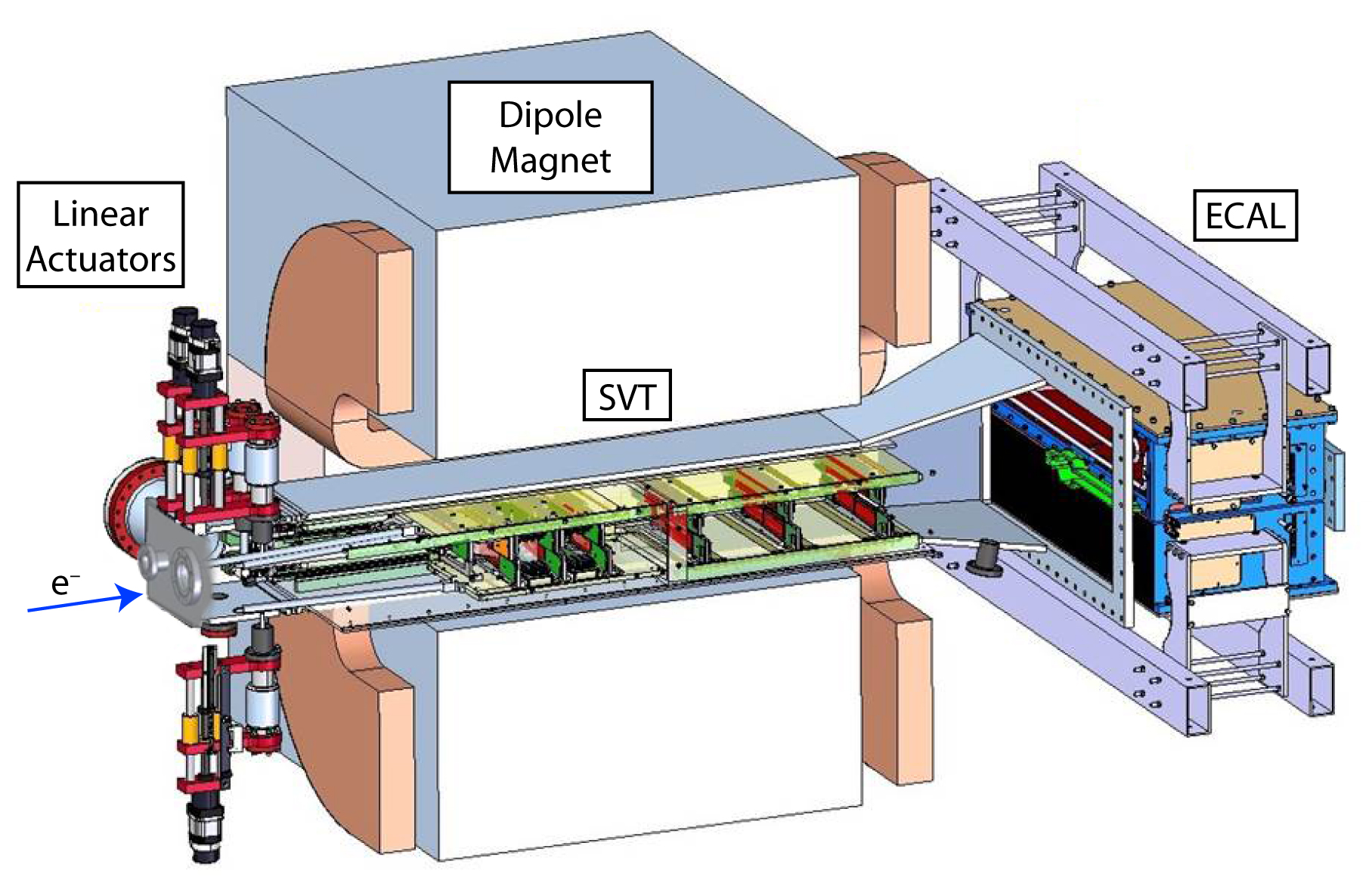}
\caption{Partial cut-out view of the HPS setup.}
\label{fig:hps_setup}
\end{center}
\end{figure*}

Fig.~\ref{fig:2h_line} shows the downstream end of the beam line in experimental Hall B, where the HPS setup is located in the alcove downstream of the CLAS spectrometer~\cite{CLAS}. The HPS setup is a forward spectrometer based on a dipole magnet, 18D36 (pole length 91.44~cm, max-field of 1.5~T) as shown in Fig.~\ref{fig:hps_setup}. The target and SVT are installed inside a large vacuum chamber within the gap of the dipole magnet. The calorimeter (ECal) is mounted behind the magnet, $134$~cm downstream of the target, and split above and below a vacuum chamber that connects the SVT vacuum chamber to the downstream beam line and the beam dump. The ECal vacuum chamber was designed to allow the innermost ECal crystals to be mounted just $20$~mm from the beam plane. It has two wider openings, one to accommodate most of the multiple scattered beam electrons and the other, the bremsstrahlung photons created in the target.

In order to transport the electron beam to the beam dump, two small dipole magnets (pole length 50~cm, max-field 1.2~T) have been installed upstream and downstream of the spectrometer, forming a three magnet chicane with zero net integrated field along the beam path. The electron beam is deflected to beam's left in the first chicane dipole. It impinges on the target, which is located at the upstream edge of the spectrometer magnet, $6.8$~cm to the left and at a horizontal  angle of $31$~mrad with respect to the original beam line. The bremsstrahlung photon beam generated in the target continues in that direction after the target, while the electron beam is bent back by the spectrometer magnet toward the second small dipole, which in turn restores the beam to its original direction in line with the dump. Behind the chicane there are two shielding walls that separate the HPS setup from the downstream tunnel. A photon beam dump (a lead cave with a tungsten insert) was installed after the first shielding wall, $\sim 7$ meters downstream of HPS. A Faraday cup cage and the Hall B electron beam dump are behind the second shielding wall.  

The HPS experiment will run with beam energies from $1$~GeV to $6.6$~GeV, and beam currents up to $500$~nA using $0.125~\%$ radiation length, or $0.25~\%$ for high energy runs, tungsten foils as targets.  
The beam parameters required to run the SVT and ECal in close proximity to the beam plane have been established using simulation and are presented in Table~\ref{tab:beam_par}. Requiring a small beam spot improves mass and vertex resolution when beam position constraints are included in the track and vertex fits. The tracking resolution at the target, which is much better in the vertical than horizontal direction, is reflected in the disparate beam size requirements in x and y. A small vertical beam size is also essential for keeping beam tails away from the Layer 1 silicon sensors.  

 \begin{table}[htb]
 \centering
 \begin{tabular}{|c|c|c|}
\hline
Parameter & Requirement &Unit \\ \hline 
Beam Energy (E) & $1$ to $6.6$& GeV \\ \hline
$\delta$E/E & $< 10^{-4}$ & \\ \hline 
Beam Current & $50$ to $500$ & nA \\ \hline
Current Stability & $\sim 5$ &\% \\ \hline 
$\sigma_x $&$< 300$& $\mu m$ \\ \hline 
$\sigma_y$&$ < 50$& $\mu m$ \\ \hline
Position Stability &$< 30$ &$\mu m$ \\ \hline
Divergence& $< 100$& $\mu rad$ \\ \hline 
Beam Halo ($> 5\sigma$) &$< 10^{-5}$& \\ \hline
 \end{tabular}
\caption{ Required beam parameters.} 
\label{tab:beam_par}
\end{table}

The Hall B beamline is well equipped to deliver high quality beams which meet these requirements. Small, stable beams have been routinely delivered for experiments in Hall B~\cite{APF_ldrp} using the CLAS detector~\cite{CLAS} where targets are positioned at the center of the experimental hall.

\subsection{Description of the Hall-B beamline}
        
The Hall B beamline is divided into two segments, the so called ``2C" line, from the Beam Switch Yard (BSY) following the beam extraction from the CEBAF accelerator to the hall proper, 
and the ``2H" line from the upstream end of the experimental hall to the beam dump in the downstream tunnel. The ``2C" part of the beamline features an achromatic double bend (dogleg) that brings beam up to the hall's beamline elevation from the BSY. For  most experiments, where the targets are located upstream of the center of the hall, instrumentation on the ``2C" line is sufficient to shape the beam profile and position it. The beam line instrumentation on the ``2H" line is then used only for monitoring beam properties.

\begin{table}[htdp]
\caption{List of elements on the Hall-B line from the beginning of the upstream tunnel to the Faraday cup in the downstream beam dump that are actively monitored and controlled by the experiment.}
\begin{center}
\begin{tabular}{|l|l|}
\hline
Description and section & L (meters) \\ \hline 
Stripline BPM, quadrupoles, & -40.2 \\ 
H/V-correctors and viewer 2C21 & \\
Wire Harp 2C21 & -38.8 \\
nA-BPM 2C21A & -37.6  \\
Stripline BPM and quadrupoles & -26.55  \\
H/V-correctors, 2C22 and 2C23 & -26.5  \\
Quadrupoles, H/V-correctors & \\ 
and viewer 2C24 & -25. \\ 
nA-BPM 2C24A & -24.5  \\
Wire Harp 2C24 & -22.0  \\
Hall-B tagger dipole & -17.6  \\ \hline
Stripline BPM 2H00 & -12.3 \\
Quadrupoles and H/V-correctors 2H00 & -11.6 \\ 
nA-BPM 2H01 & -8.0  \\ \hline
Center of the hall & 0  \\ \hline
Stripline BPM 2H02 & 13.5  \\ 
SVT collimator & 14.1  \\
Wire harp 2H02A & 14.8  \\
Chicane dipole 1 & 15.3 \\
HPS target & 17.0  \\
Spectrometer dipole & 17.5 \\
Chicane dipole 2 & 19.7 \\
Beam viewer 2H04 & 24.0  \\
Dump, Faraday cup & 27.0  \\
\hline
\end{tabular}
\end{center}
\label{tab:elements}
\end{table}%

As seen in Fig.~\ref{fig:2h_line}, the HPS setup is located at the downstream end of the 2H beam line. The HPS target is about $17$ meters downstream of the nominal Hall B center. In order to deliver a small sized beam to HPS with the required position stability, an additional set of quadrupoles and corrector dipoles and new beam diagnostic elements have been added to the beamline. This design was optimized using the beam transport software package ELEGANT~\cite{elegant}. The beamline optimizations have incorporated the 12 GeV CEBAF machine parameters and targeted the HPS beam size requirements~\cite{hpsbeam}. The whole framework has been validated experimentally using the 6 GeV CEBAF~\cite{elvalid}. The optimization was done for three beam energies, $1.1$~GeV, $2.2$~GeV, and $6.6$~GeV.  A two-quadrupole girder based on the existing design of the CEBAF arc recombiner can deliver the desired beams when the girder is placed about $12$ meters upstream of the hall center.

The  devices which are actively used to control the beam in the hall and  monitored by the experiment are listed in Table~\ref{tab:elements}. The list starts from the shielding wall that separates the BSY and the Hall B upstream tunnel (this is where the beam gets to the Hall B beamline elevation). The first column in the table is the description of the element and the name used to identify it, and the second column gives its position relative to the geometrical center of the hall. The critical elements for shaping the beam profile on the HPS target are the two quadrupoles on the 2H00 girder. The required beam position stability was achieved by feeding back the readings of two stripline Beam Position Monitors (BPM)~\cite{striplineBPM}, 2H00 and 2H02, to control the horizontal and vertical correctors on the 2C22, 2C23, and 2H00 girders.   

\begin{figure*}[t]
\begin{center}
\includegraphics[width=0.75\textwidth]{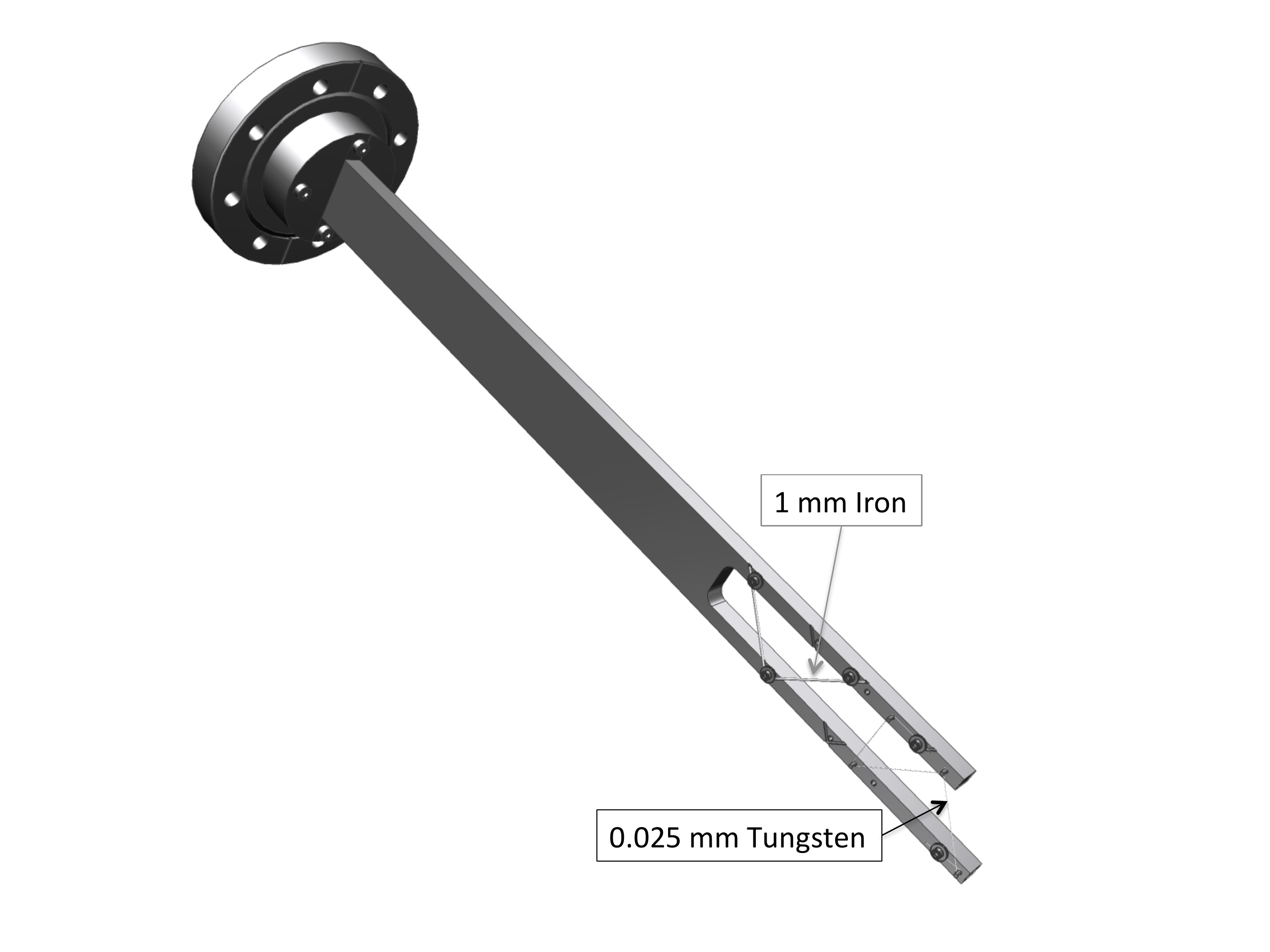}
\caption{2H02A Wire Harp. The wire frame is moved into the beam at $45^\circ$. }
\label{fig:2h02harp}
\end{center}
\end{figure*}

The beam profile and position are routinely measured by the wire scanner ``harps'' located strategically along the beam line. In particular, the wire harp ``2H02A'' (shown in Fig.~\ref{fig:2h02harp}), located only 2.2~m upstream of the target, played an important role in confirming the beam profile and position required for the experiment. This harp measures the beam profile and its projected position along x-, y-, and 45$^\circ$ axes. The beam profile and position are extracted by correlating the halo monitor response (distribution of photomultiplier (PMT) in proximity of the beam line) with the wire position as the wire moves through the beam. Multiple wires are mounted on a {\it fork} at $\pm 45^\circ $ and $90^\circ$ angles with respect to the fork and the fork axis of motion is at 45$^\circ$ with respect to the horizontal plane. The count rates from these halo monitors are also used as inputs to the machine fast shutdown system (FSD)~\cite{fsd} as described below. In addition to the actively monitored and controlled devices, a $10$~mm thick tungsten collimator was installed $3.1$ meters upstream of the HPS target for SVT protection as shown in Fig. \ref{fig:2h_line}. It had three  rectangular holes of various sizes. For most data taking during the 2016 run the $2.82$~mm$\times 10$~mm hole was used.  

The HPS experiment has run with beam energies of 1.05 and 2.3~GeV, and plans to run up to 6.6 GeV with beam currents up to 500~nA. The chicane magnet settings are scaled with beam energy in order to maintain the beam trajectory. HPS uses a tungsten foil target, 0.125~\% of a radiation length thick at the lower energies, or 0.25~\% of a radiation length at energies of 4.4 and 6.6~GeV. The foil is supported on the sides and top by a frame that has been kept very thin. This is to prevent a significant radiation dose to the silicon detectors in case the beam would be accidentally deflected off the target foil by, for example, an upstream magnet tripping off. The target, supported on a cantilever and moved vertically by a linear actuator located outside the vacuum chamber and spectrometer magnet, can be lowered on to the beam line without interrupting the beam operation.

\subsection{SVT Mover and Wire scanner}
\begin{figure}[htpb]
\begin{center}
\includegraphics[width=1.0\textwidth]{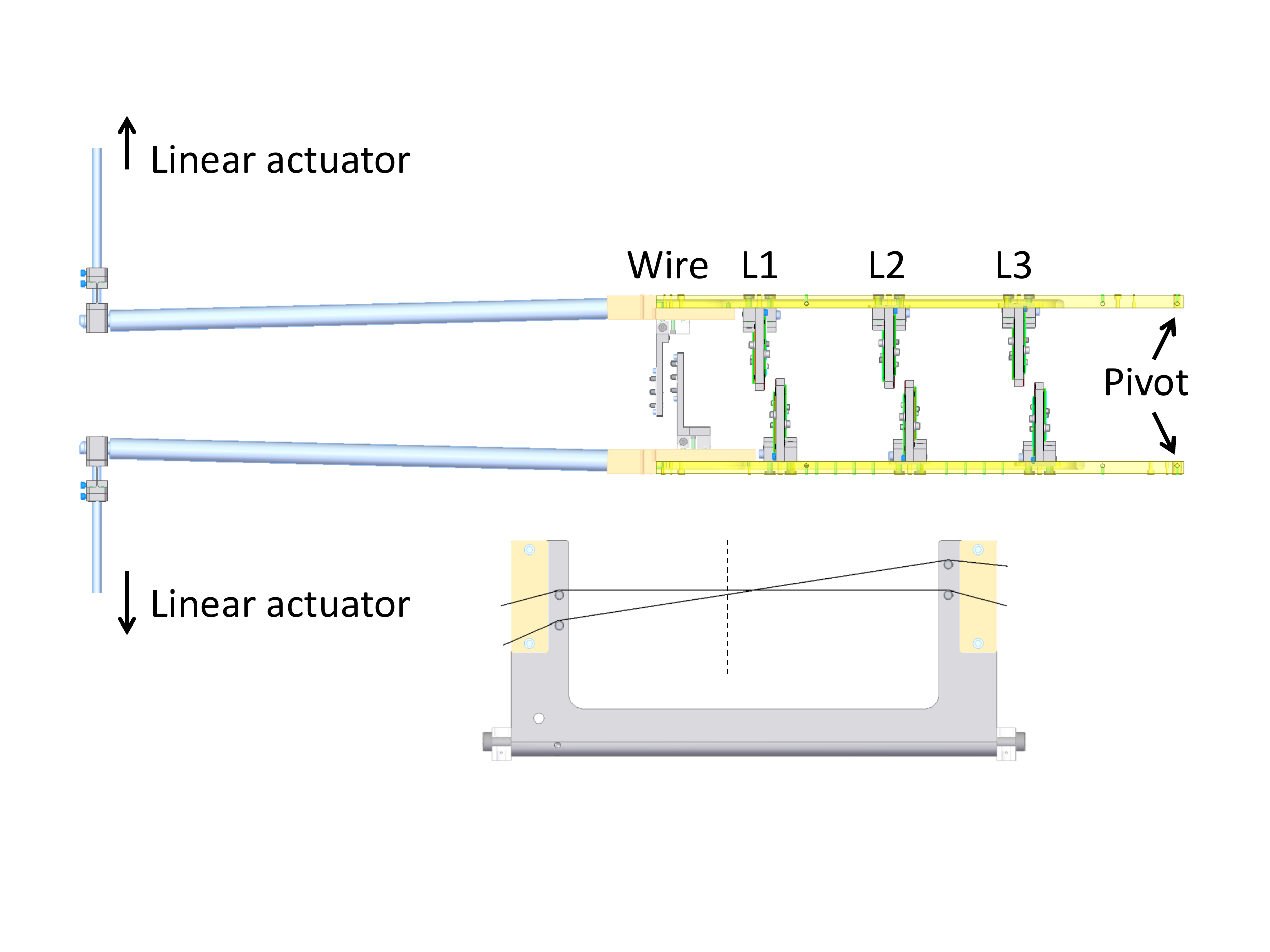}
\caption{Elevation view of the SVT motion system and beam's eye view of the SVT wire scanner with dashed line indicating the nominal horizontal beam position.}
\label{fig:motion}
\end{center}
\end{figure}

In order to bring the SVT to within 500 $\muup$m of the beam, the top and bottom SVT layers 1-3 are mounted on movable support plates as shown in Fig.~\ref{fig:motion}. Each support plate holds three sensor layers comprised of both axial and stereo sensors, and a wire frame. The plate is supported by two pivots on a downstream ''C-support'' and  connected to a lever extending upstream to a precision linear actuator which can move it up and down. In the nominal run position, the support plates are positioned parallel to the beam and the physical edge of the sensor is 0.5 mm (L1), 2.0 mm (L2), and 3.5 mm (L3) from the beam. The SVT can be retracted during beam tuning or when the beam is very unstable. At the fully retracted position, the layer 1 sensor edge is 8 mm from the beam.

Also shown in Fig.~\ref{fig:motion} is a beam's eye view of the bottom wire frame which holds two wires and is also directly mounted on the SVT support plate. The horizontal wire is 20 $\muup$m diameter gold-plated tungsten. The angled wire is 30 $\muup$m diameter gold-plated tungsten and placed at 8.9 degrees to the horizontal wire. While the wires are scanned across the beam using the linear actuator, the downstream halo counter rates are recorded at each position. Since the horizontal wire position had been surveyed in the lab with respect to the nominal SVT layer 1 sensor edge with 50-100 $\muup$m precision, the SVT wire scan is used to determine the position of the beam relative to the nominal center of the SVT coordinate system. After moving the beam to the desired position in the SVT coordinates, and moving the support plates to their nominal positions, the silicon sensors in each of the layers are then correctly positioned  with respect to the beam. The distance between the horizontal and angled wires is used to measure the horizontal beam position relative to the SVT coordinate system with about 300 $\muup$m uncertainty. If necessary, the beam is moved horizontally, so that it is properly positioned with respect to the SVT. 

\section{Beamline performance}
\label{performance}

The HPS engineering runs in 2015 and 2016 shared time in Hall B 
with the CLAS detector upgrade project. Beams for HPS were available only after regular work hours, on evenings and weekends. 
This  arrangement required re-establishing high quality beams daily. Such operations relied heavily on the 
reproducibility of the beam parameters from CEBAF, the stability of beamline magnet settings, the performance of beam position and halo monitors, 
and the efficiency of beam set up procedures.

\subsection{Establishing beam for physics}

\begin{figure}[htpb]
\begin{center}
\includegraphics[width=0.5\textwidth]{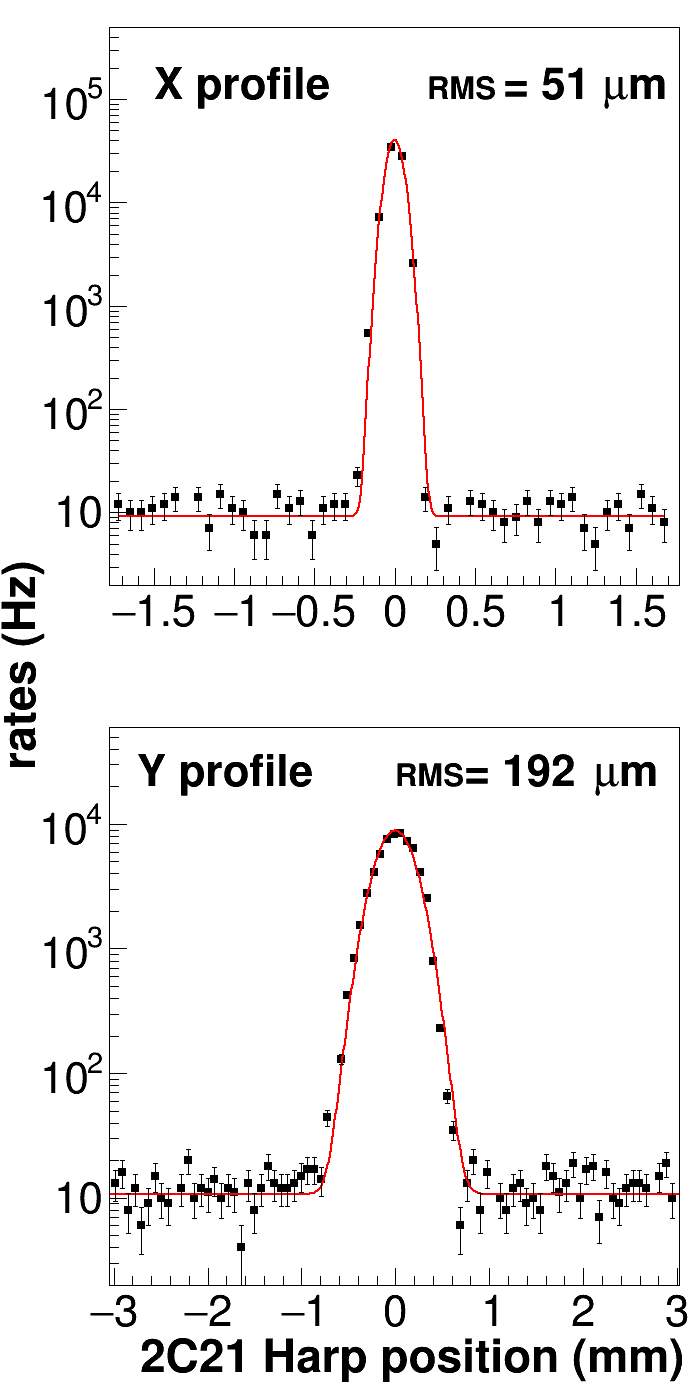}
\caption{Beam profile measurement with wire harp 2C21 during the 2.3 GeV run.}
\label{fig:2c21}
\end{center}
\end{figure}

Establishing production quality electron beam for experiments in Hall B is a two step process. The initial tune is done at low current by deflecting the beam down to an intermediate dump with the Hall 
B tagged photon spectrometer dipole magnet~\cite{tagger} and establishing the required beam profile and positions using wire harps and nanoamp (nA) BPMs~\cite{nA_BPM} in the upstream tunnel at the 2C21 and 2C24 girders (see Table~\ref{tab:elements}). An example of a beam profile at 2C21 from the $2.3$~GeV run is shown in Fig.~\ref{fig:2c21}, confirming that the beam was delivered to the hall with required beam profile.

\begin{figure}[htpb]
\begin{center}
\includegraphics[width=0.5\textwidth]{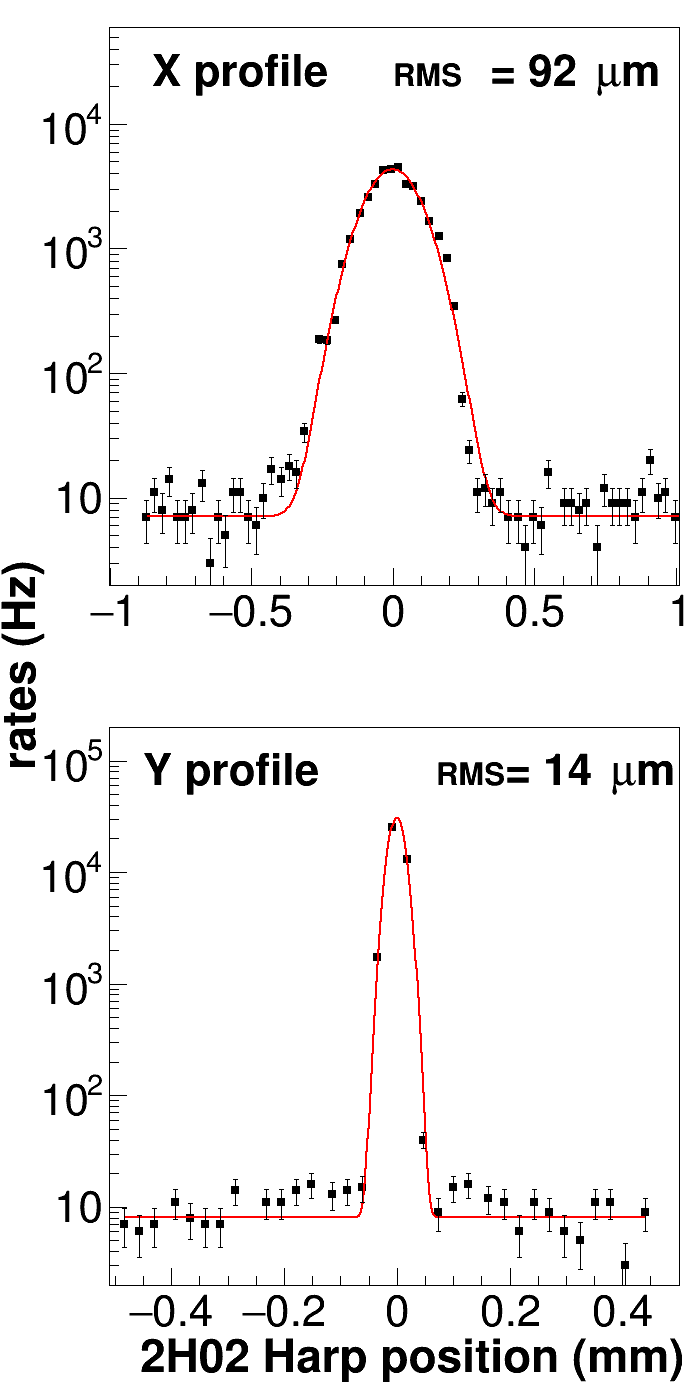}
\caption{Beam profile measurement with wire harp 2H02A during the 2.3 GeV run.}
\label{fig:2h02a}
\end{center}
\end{figure}

 In the second step, after degaussing and turning off the tagged photon spectrometer dipole, 
the beam is sent straight to the electron dump at the end of the Hall B beamline.  Tuning and positioning the downstream 
beam profile is done using the 3-wire harp ``2H02A'' mounted about $2.2$ meters upstream of the HPS target.  An example of (X,Y) beam profile from the $2.3$~GeV run is shown in Fig.~\ref{fig:2h02a}. Beam sizes and positions were initially measured
on the harp and two stripline BPMs, 
2H00 and 2H02. The overall beam position was then moved as needed to center the beam on the SVT coordinate system using the 
SVT wire scanner (see above). The SVT protection collimator was aligned with respect to the beam and remained inserted 
at all times. The downstream beam viewer at 2H04, located just before a Faraday cup, had three interchangeable screens, 
a Chromox disk, a YAG crystal, and an Optical Transition Radiation foil. It was used to ensure clean beam transport through the SVT collimator 
and the chicane and to monitor beam stability during data taking.
     
\begin{figure}[htpb]
\begin{center}
\includegraphics[width=1.0\textwidth]{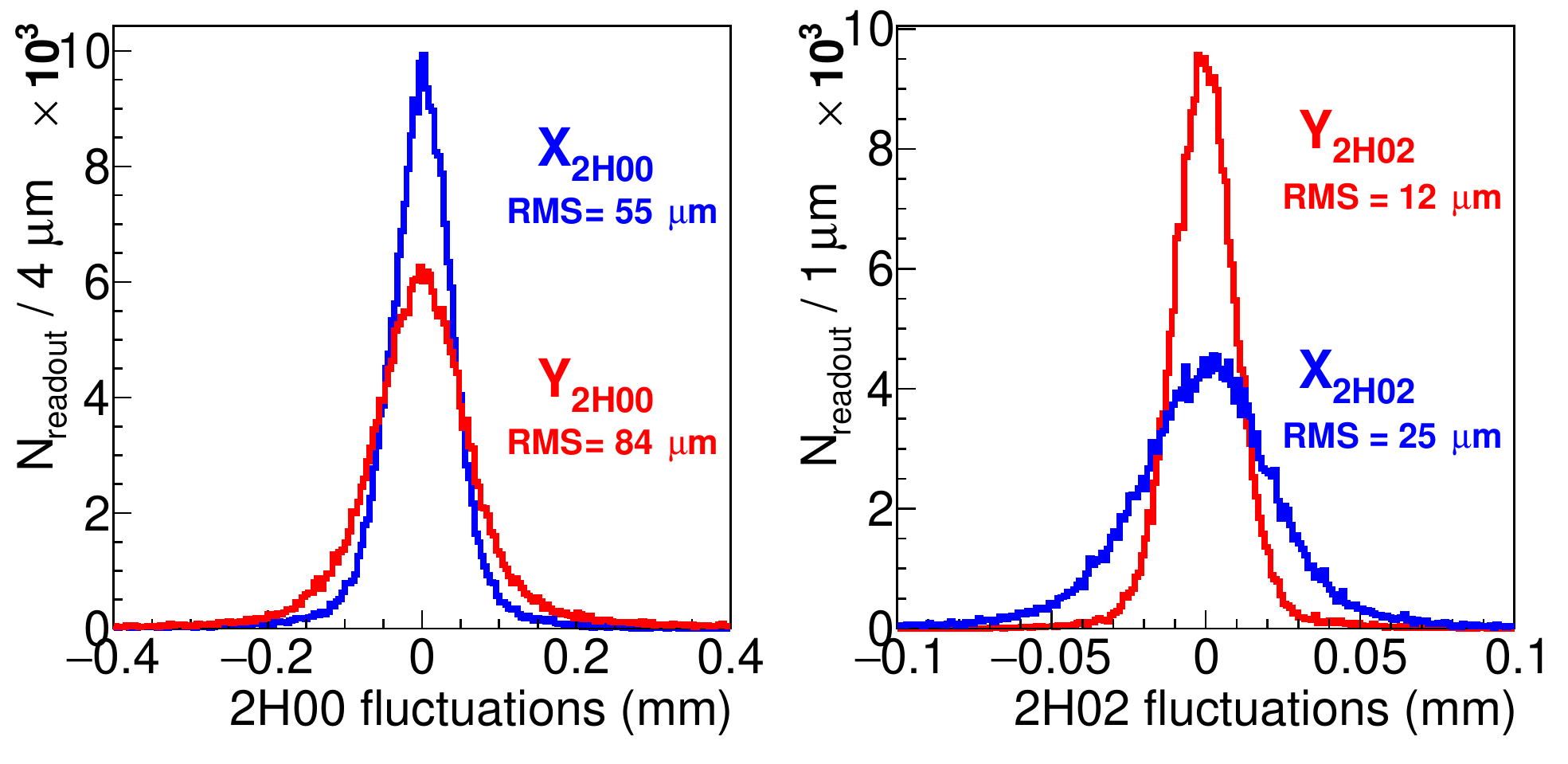}
\caption{Distributions of beam positions at BPM 2H00 and 
2H02A during the 2.3 GeV run.}
\label{fig:bpm}
\end{center}
\end{figure}

After a high quality beam was established and properly aligned, 
the beam orbit lock system was engaged. This system uses position readings from the two stripline BPMs to regulate 
currents in the horizontal and vertical corrector dipoles to minimize beam motion at the target. The response time 
for the beam orbit corrections is of order one second and is determined by the readout bandwidth of the stripline BPMs 
at the operating beam currents. The position accuracy of the BPMs depends on the beam current and the readout speed. The data 
from the BPMs were read out and archived at a $1$ Hz rate. In Fig.~\ref{fig:bpm} the variations of ($x,y$) positions from 
the set points on these two BMPs during the entire data taking period of 2016 run are shown. The variations on the upstream 
BPM, labeled 2H00, are $RMS\simeq 55~\muup$m and $\simeq 84~\muup$m for $x$ and $y$ positions, respectively. The variations 
on the downstream BPM, 2H02A, are less than $30 ~\muup$m in both $x$ and $y$ and are within the resolution of the stripline BPMs. Similar 
position stability was present during the 2015 run. Based on these data and the data from harp scans where the beam profile 
is sampled in less than a second (the halo counter scaler readout rate is $\sim 15$ Hz), we conclude that any beam motion at the 
target at a frequency of a few Hz is smaller than the beam width.  Based on the BPM data the vertical beam angle drift 
is less than $0.4 ~\muup$rad as the distance between the two BPMs was $25.5$ meters. The observed beam angle variations 
have negligible impact on the beam's trajectory through the SVT.

Once a stable beam at the target is established, the chicane magnets are 
energized and the HPS target is inserted. The final step in establishing the production running conditions is setting 
limits on the halo counter rates for the beam Fast Shut Down (FSD) system. If the beam moves close to obstacles, e.g. 
collimator walls or silicon detector edges, count rates on the beam halo monitors will increase. The threshold for the 
FSD trip  
was set on the sum of the rates of the four halo monitors mounted right before and after the HPS target. These rates 
were essentially constant when beam conditions were stable. Most HPS data taking took place with FSD limits set to trip 
the beam if the sum of the halo counter rates exceeded its mean value (for a fixed integration time) by $5.5\;\sigma$.  The integration 
time windows used during the 2015 and 2016 runs were $5$ ms and $1$ ms, respectively. The trip point corresponds to one false FSD trip 
every $16$ hours if the fluctuations are purely statistical.

While the FSD system could prevent errant beam hitting the SVT within a milli-second time frame, it could not react to faster beam motions. 
The SVT protection collimator fully protected the active region of the silicon detector in Layer 1, but its 
protection did not extend into the 1 mm wide guard ring surrounding the silicon active area. 
It was therefore important to understand if there is any sizable beam motion at a much faster rate, outside of FSD and BPM 
response times, that might irradiate the guard ring. A system capable of sampling and storing halo monitor counts at tens of 
kHz rate was deployed to study possible fast beam motion. This system, which was also used to test the FSD system and to 
study beam motion during beam trips and beam restoration, is described below.

\subsection{Beam alignment relative to SVT}

\begin{figure}[tb]
\begin{center}
\includegraphics[width=1.0\textwidth]{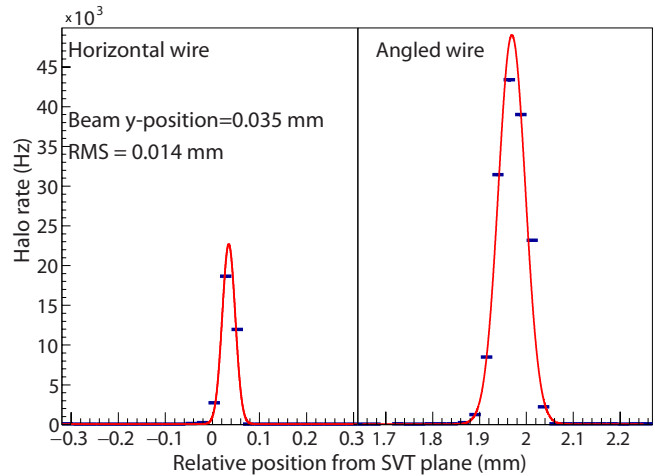}
\caption{SVT wire scan.}
\label{fig:wirescan}
\end{center}
\end{figure}

When the beam is delivered for the first time to HPS, or delivered after a long down time, an SVT wire scan is performed to check 
that the beam is still aligned to the SVT coordinate system. If the beam was not well-aligned, it was re-centered. 
Fig.~\ref{fig:wirescan} shows the halo counter rate as a function of the linear actuator position for the top SVT wire scanner. The first 
peak is from the horizontal wire and the rate difference is due to the wire size difference. After fitting each peak to a Gaussian, 
the vertical and horizontal beam positions relative to the SVT coordinate system are measured. When the beam position was off 
more than 100 $\muup$m, the upstream corrector magnets were used to move the beam, and the SVT wire scan was repeated to confirm 
the beam movement. The wire scan shown in Fig.~\ref{fig:wirescan} resulted in the vertical beam position of 35 $\muup$m above and 
the horizontal beam position of 99 $\muup$m left of the SVT coordinate system, well within tolerances for beam placement. The 
vertical beam size was measured from the horizontal wire scan to be 14 $\muup$m, which was consistent with the measurement by the 
2H02 harp after accounting for the fact that the beam is being focused to the target location, and the harp is about 2 meters 
upstream. Once the beam position was established, the SVT wire scan was 
performed only sparingly since the orbit lock system 
could maintain the beam position well within $50\;\muup$m.

\subsection{Beam motion studies}

Several studies have been done to understand beam position stability beyond those possible 
with the  standard control and monitoring system. Signals from halo monitors downstream of the HPS target were fed to a 
Struck SIS3800 scaler VME module read out by EPICS Input-Output Controller (IOC) running on a Motorola MVME5500 CPU board. 
This application allowed us to latch the scalers for time intervals as short as 15 microseconds and to write them into ROOT~\cite{root} 
files for offline analysis. The following studies were done using this system to look for short term beam motions.

\subsubsection{Fast beam motion}

To get reasonable halo counter statistics in a 15 microsecond window, the beam must be parked 
close to a solid object so its tails can produce high backgrounds. For the first study, the $1$~mm thick iron wire strung 
horizontally on 2H02A harp (Fig.~\ref{fig:2h02harp}) was positioned close to the beam core to get sufficient rates. A change of rate indicates beam motion 
towards or away from the wire. Both wire positions, above and below the beam, were studied, to be sensitive to possible fast beam 
motions in either direction.


\begin{figure*}[htb]
  \centering
  \includegraphics[width=0.8\textwidth]{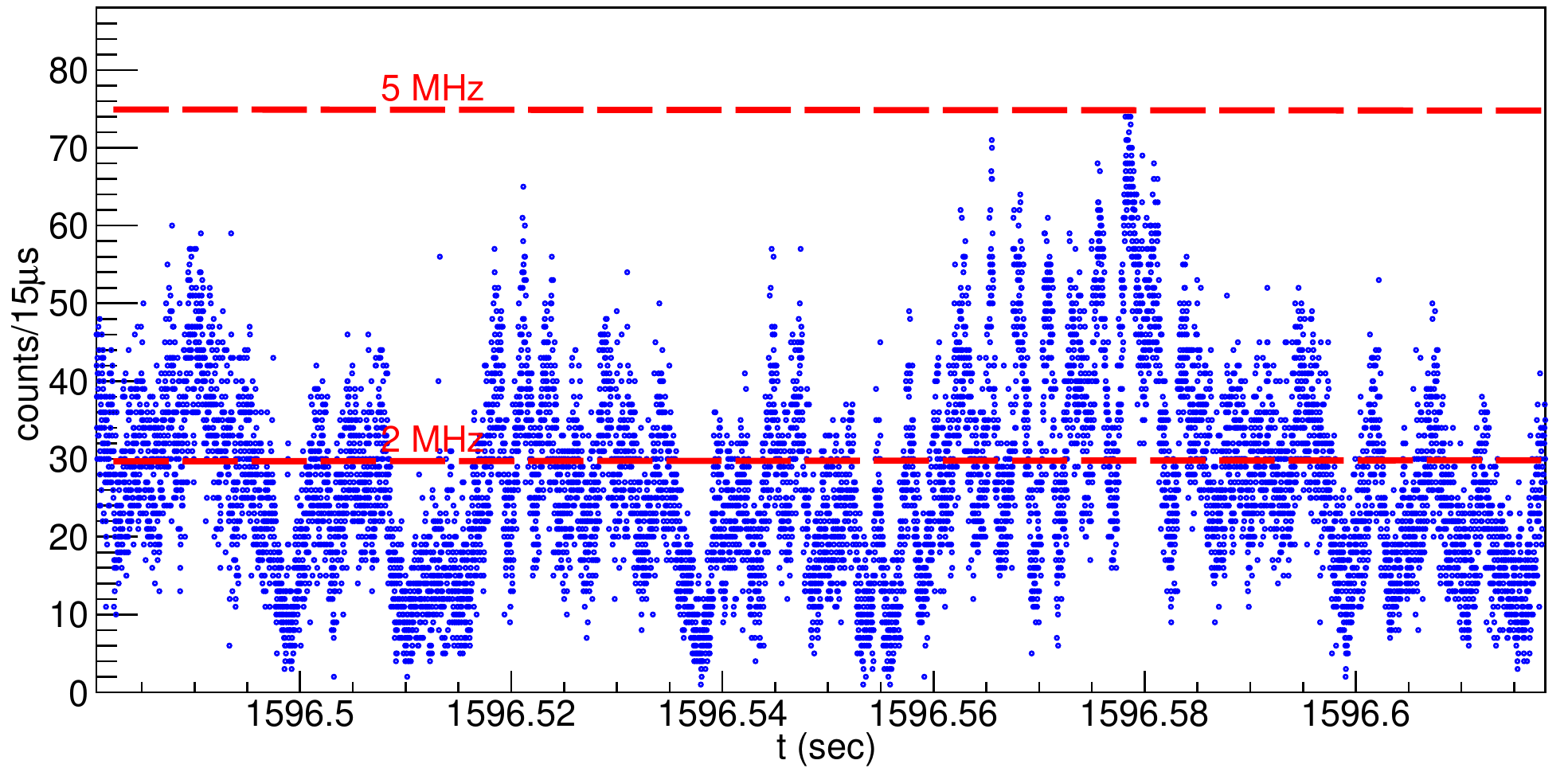}
  \caption{Oscillations of 
rates on the halo monitors in the $140$~ms time interval.}
  \label{fig:Struck_Oscillations}
\end{figure*}


Fig.~\ref{fig:Struck_Oscillations}~ shows a $140$ ms snapshot of the scaler readout. With $100$ nA beam 
current, the average rate was $30$ counts per $15~\muup$s, or $2$ MHz. The maximum rate was about $75$ counts 
per $15~\muup$s or $5$ MHz. Since such rapid rate fluctuations were not observed on the same monitors when the 
HPS target was inserted, the observed fluctuations are attributed to oscillations in the beam position.

 
\begin{figure}[!htb]
  \centering
  \includegraphics[width=0.5\textwidth]{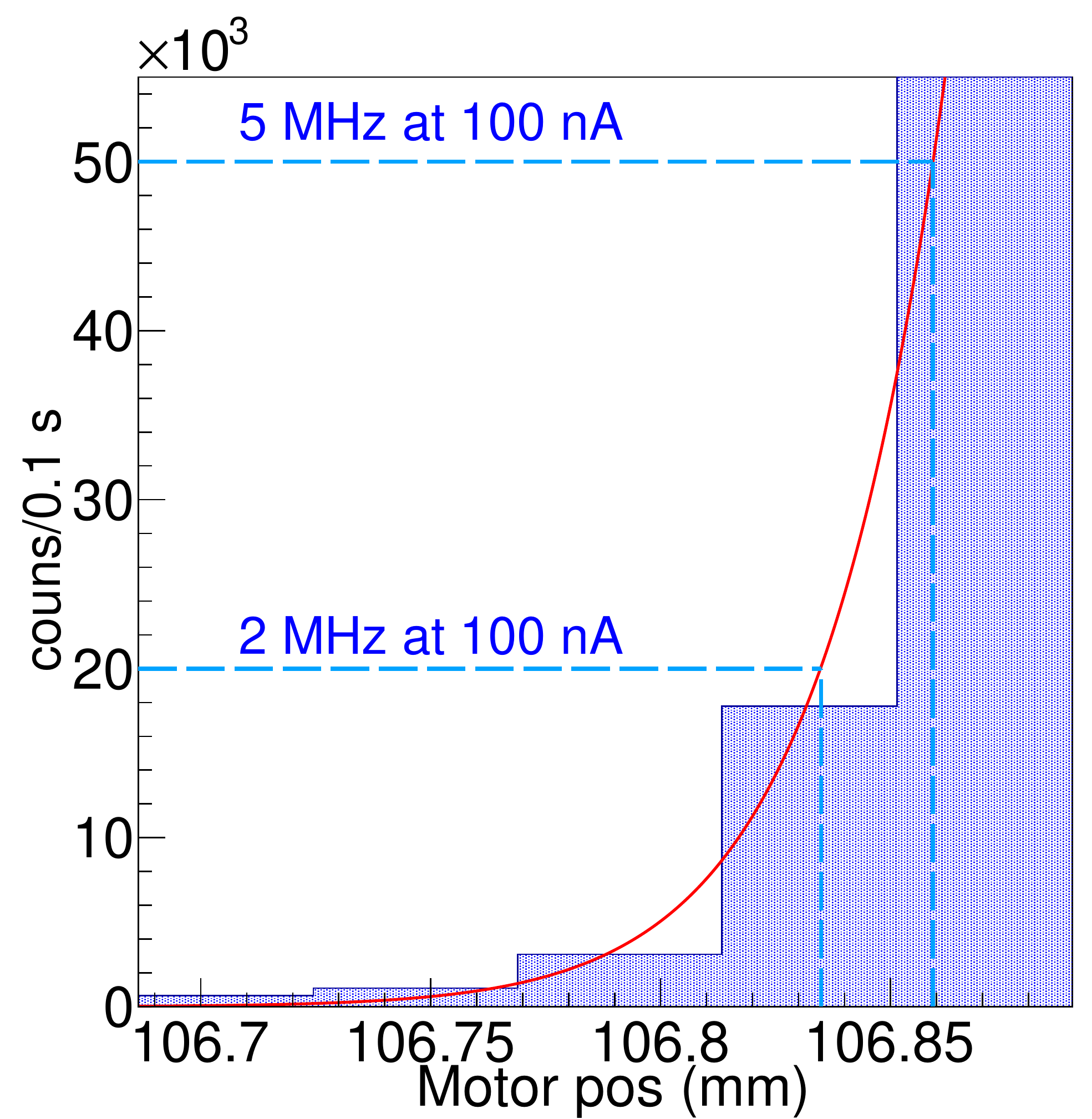}
  \caption{Counts on the halo monitors as a function of harp motor position. The red solid line is the Gaussian fit to the whole beam profile. The two blue dashed vertical lines show the positions of the wire where rates are $2$ MHz and $5$ MHz.}
  \label{fig:HarpScan_ThickWire}
 \end{figure}


To estimate how large 
these beam motions are, the same halo rates were measured every 0.1 second while the wire was passing through a $10$ nA 
beam at a speed of $0.5$ mm/sec, while the full beam vertical profile was measured.  Fig.~\ref{fig:HarpScan_ThickWire}~ shows 
a small portion of that scan corresponding to positions between the beam and the wire at which the rate fluctuation data shown 
above was taken. The red line on the figure corresponds to the Gaussian fit to the beam profile. The motor positions  
corresponding to $2$ MHz and $5$ MHz count rates at $100$ nA beam current (after correction for the readout speed and 
the beam current) are shown with blue  dashed lines. The difference between these two positions, after accounting for 
the fact that the harp moves at 45 degrees with respect to the vertical axis, is $\approx 17.7~\muup$m. So vertical beam 
oscillations of just $18~\muup$m will account for the observed rate variations, and this is much smaller than the vertical 
beam size of $\sim 30~\muup$m. Fast beam motion is not a problem for HPS. 


\subsubsection{Beam motion during beam trips}

There are various causes for beam trips:  beam loss in the machine itself; beam loss in the delivery lines which cause rate increases on beam loss or halo monitors; RF system trips or loss of RF to an accelerating cavity. The average beam trip rates were $5$ and $10$ per hour during HPS runs in 2015 and 2016, respectively. Although care was taken to develop optics with no vertical dispersion in the HPS region to avoid instabilities due to RF fluctuations, beam motion preceding a beam trip could conceivably result in a beam strike on the exposed regions of the SVT before beam termination. The time it takes to shut the beam off is expressed as: 

\begin{eqnarray}
t=27 \muup \textrm{s}+N_{pass}\cdot 4.2 \muup \textrm{s}/pass+\textrm{Detection time}  
\end{eqnarray}
where $27~\muup$s is the 
time it takes to shut the beam at the injector and $4.2~\muup$s/pass\footnote{A pass is a complete turn in the CEBAF accelerator.} is the time it takes to clear the machine. The ``Detection Time" depends on the source of the trip and can take milliseconds, e.g for the HPS halo monitor FSD system 
that time is $1$ ms. If there is a large beam motion towards the SVT or other obstacles (collimator or beam pipes), this should 
be captured in the halo monitor rates when they are read out rapidly. Using the Struck scaler 
setup  a couple of dozen beam trips and beam recovery periods have been recorded. There were no rate increases during the beam recovery,
indicating no appreciable beam motion. In $30~\%$ of the beam trips, some rate increase was observed within $200~\muup$s prior to beam shut off.
This could have been due to beam motion at the SVT or to higher beam backgrounds passing through the entire beam line. Our measurement
does not differentiate between these possibilities. Since many of these instances of increased halo counter rates coincided with increased upstream 
halo counter rates, it is likely that beam motion was not to blame in these cases.


%



\subsubsection{Test of the FSD system}

The CEBAF Fast Shut Down (FSD) system provides permissive signals to the injector gun. 
If the permissive is removed, the gun shuts 
off within 27 $\muup$s. 
The sum of the HPS halo counter rates was used for one of the inputs to the FSD system. 


\begin{figure}[!htb]
 \centering
 \includegraphics[width=1.0\textwidth]{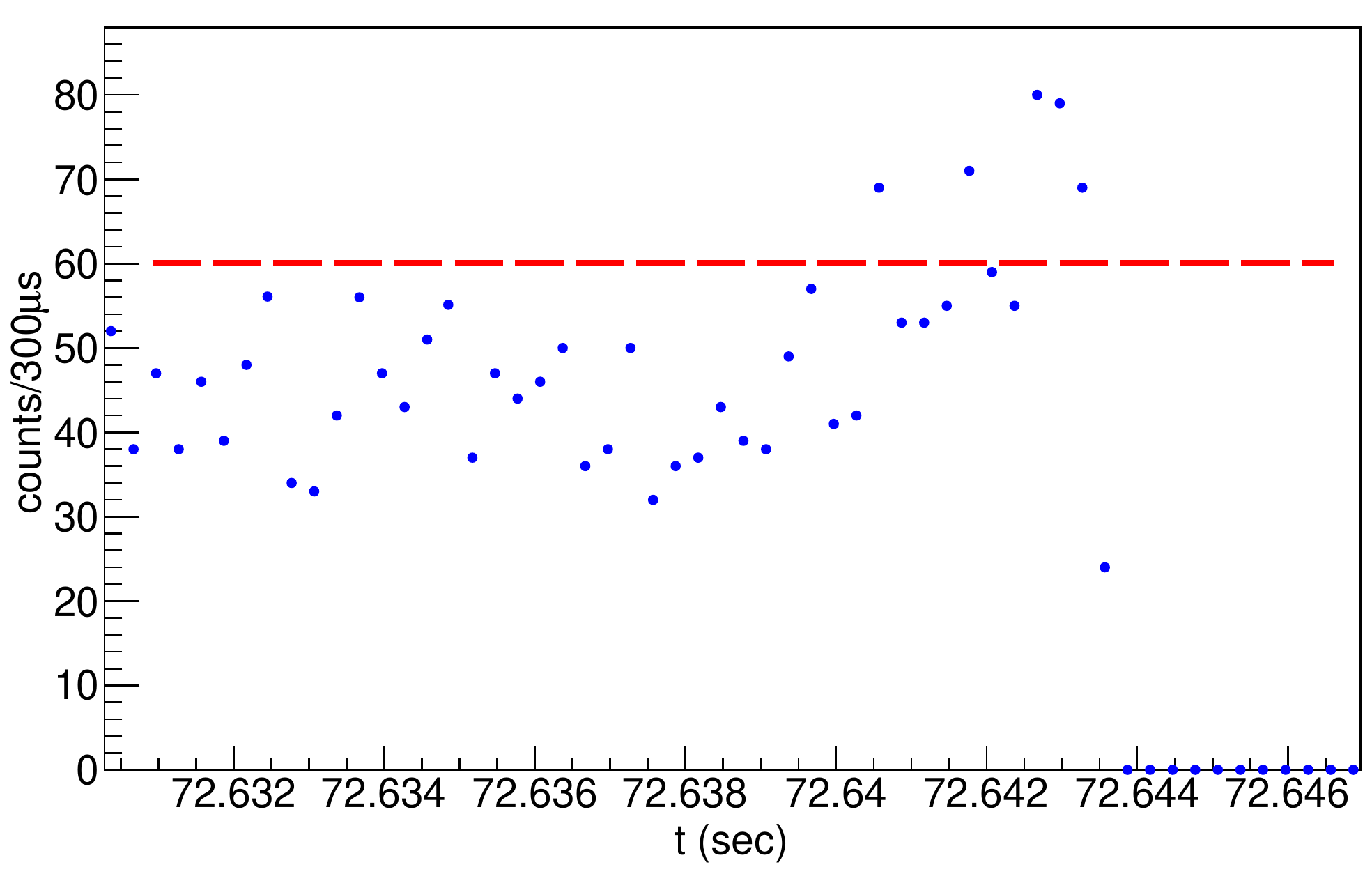}
 \caption{HPS halo counter rates 
as a function of time. The FSD integration time is $1\; ms$ and the the beam
current is 200 nA. Each point represents integrated 
counts over a $300\;\muup s$ time interval.}
 \label{fig:FSD_Test}
\end{figure}


After setting up production running, the halo counter rates were increased by running the 2H02A harp wire through the beam, simulating the condition where the beam tail is hitting an obstacle. The system was tested for $1$~ms, $5$~ms, and $10$~ms integration times. 
In Fig.~\ref{fig:FSD_Test} one of the measurements with FSD integration time set to $1$~ms is shown. Each point on the graph 
is the integrated counts over $300\;\muup s$. The red line on the graph is the trip threshold, $60$ counts per $300~\muup$s or $200$ kHz. 
As one can see there are four cases when rates go above the red line, three of which have only one point ($300~\muup$s) above the threshold. 
The fourth one has three points ($3\times 300~\muup$s) above the red line after which beam tripped as one would expect for trip 
integration time of $1$~ms. The same tests with $5$~ms and $10$~ms integration times gave consistent results. 



\subsection{Beam halo}

\begin{figure}[!htb]
 \centering
 \includegraphics[width=1.0\textwidth]{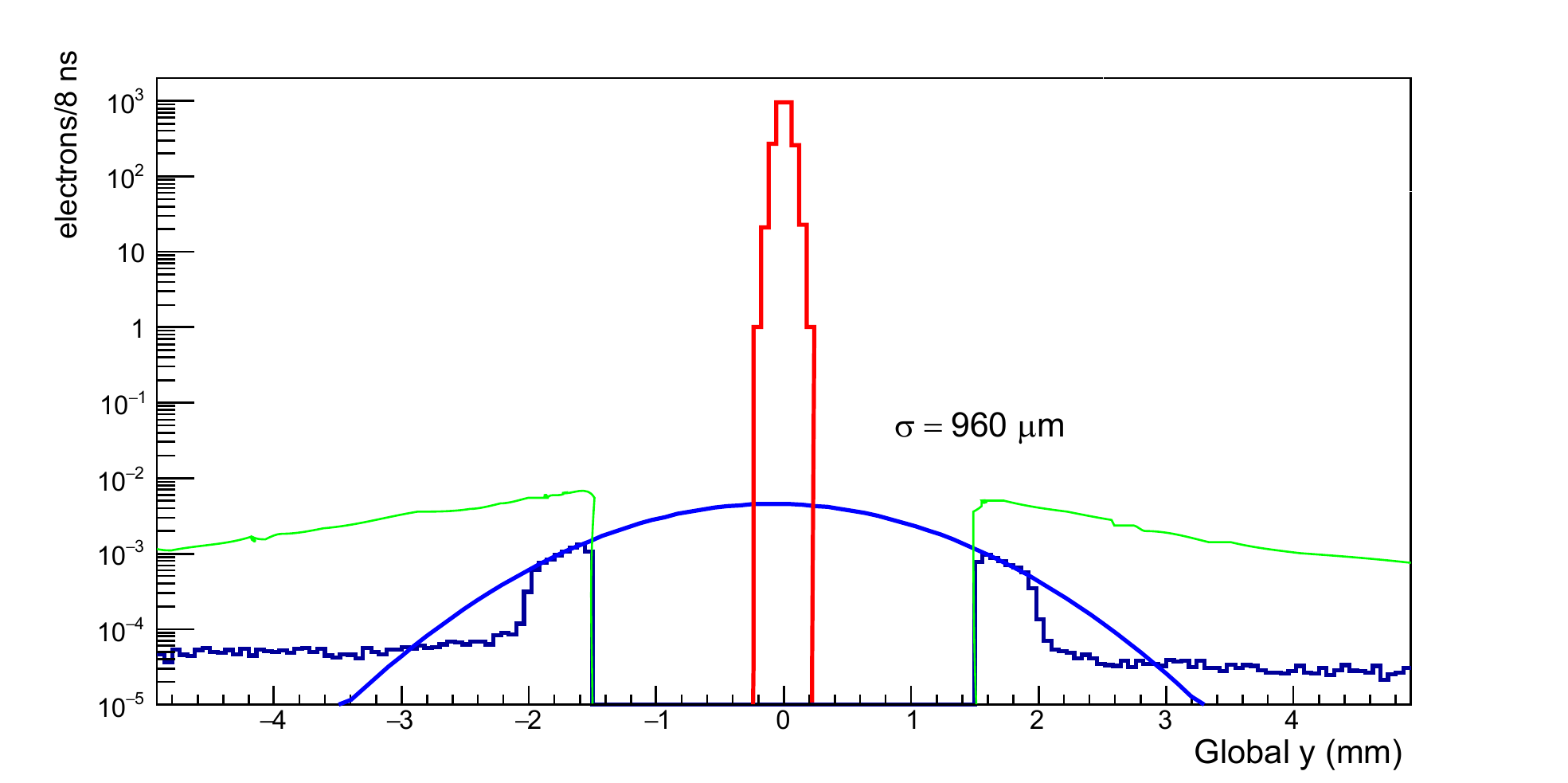}
 \caption{SVT axial layer 1 occupancy during data taking run (green) and no target run (blue). The histogram in red is nominal 50 nA Gaussian core and the blue curve is a Gaussian fit to the halo with $\sigma$ of 960 $\muup$m.}
 \label{fig:halo}
\end{figure}

HPS is very sensitive to excess beam halo. As the active area of the silicon sensor at layer 1 starts at 1.5 mm, we require that any beam halo extending beyond 1.5 mm does not contribute significantly to the sensor occupancy.
Fig.~\ref{fig:halo} shows the SVT axial layer 1 occupancy during 2015 data taking at 50 nA and a special run without the target. The SVT protection collimator with 4 mm $\times$ 10 mm hole was used. The 0.7~\% occupancy during data taking was mostly due to Coulomb scatterings in the target and was consistent with expectation. Extra hits from the beam halo were observed in the first ten channels when the target was removed. However, the extra hits were less than 20~\% of the nominal occupancies. The halo was consistent with a Gaussian distribution with $\sigma$ = 1 mm at the intensity level of $10^{-5}$ of the core beam and was consistent with the large dynamic range harp scan~\cite{APF_ldrp}. The sharp drop in the halo hits beyond 2 mm is due to the collimator, and no extra hits from the beam halo were observed during the 2016 run as the vertical size of the collimator hole was reduced to 2.82 mm.

\section{Summary}
The HPS experiment took data successfully at 1.05 GeV and 2.3 GeV beam energies with the silicon sensor edge at only 500~$\muup$m from the beam. High quality beam was delivered with the vertical beam size of 14~$\muup$m and the beam halo as small as 10$^{-5}$. The vertical beam position was maintained within 20~$\muup$m throughout the run  by the beam feedback system. The fast shut down system worked in protecting the silicon sensors from errant beam exposure.  

\section{Acknowledgements}
The authors are grateful for the outstanding efforts by the staff of the Accelerator Division and the Hall B engineering group at Jefferson Lab during the installation and running of the experiment. We also thank the HPS collaboration for taking high quality data.
This material is based upon work supported by the U.S. Department of Energy, Office of Science, Office of Nuclear Physics under contract DE-AC05-06OR23177, and Office of High Energy Physics under contract DE-AC02-76SF00515.




\bibliographystyle{model1-num-names}

\begin{thebibliography}{00}

\bibitem{HPS_prop} A. Grillo, et al. (HPS Collaboration), {\it HPS Proposal for 2014-2015 Run}, 2014. \url{http://confluence.slac.stanford.edu/download/attachments/86676777/hps_2014.pdf}
\bibitem{kinetic_mixing} B. Holdom, Phys. Lett. B 166, 196 (1986).
\bibitem{CLAS} B. Mecking, et al., Nucl. Inst. and Meth. A 503, 513 (2003).
\bibitem{APF_ldrp} A. Freyberger, Proceedings of the DIPAC 2005, Lyon France, pp 12-16; JLAB-ACC-05-318 (2005).
\bibitem{elegant} M. Borland, A Flexible SDDS-Compliant Code for Accelerator Simulation, ANL, Argonne,
IL 60439, USA
\bibitem{hpsbeam} A. Freyberger, Proceedings of IPAC2015, Richmond VA, ISBN 978-3-95450-168-7.
\bibitem{elvalid}HPS update to PAC39, \\ 
\url{https://www.jlab.org/exp_prog/proposals/12/C12-11-006.pdf}
\bibitem{striplineBPM}P. Evtushenko, A. Buchner, H. Buttig, P. Michael, B.
Wustmann, K. Jordan, {\it Stripline BPM Monitors for ELBE}, Proc. DIPAC 2001, ESRF, Grenoble, France.
\bibitem{fsd} J. Yan and K. Mahoney, Proceedings of ICALEPC2009, Kobe Japan, pp 582-584.
\bibitem{tagger} D. Sober et al., Nucl. Inst. and Meth. A 440, 263 (2000).
\bibitem{nA_BPM} M. Piller, et al., JLAB-ACC-99-30 (1998).
\bibitem{root} R. Brun and F. Rademakers, {\it ROOT - An Object Oriented Data Analysis Framework,} Proceedings AIHENP'96 Workshop, Lausanne, September 1996. \url{http://root.cern.ch/}
\end{thebibliography}



\end{document}